\documentclass[aps,prl,reprint,showpacs]{revtex4-1} 
\usepackage{epsfig,amsmath,amssymb}
\usepackage{graphicx}
\usepackage[english]{babel}

\begin{document}

\title{ On the quantum-field description of many-particle \\ Fermi
systems with spontaneously broken symmetry}
\author{Yu.\,M. Poluektov}
\email{yuripoluektov@kipt.kharkov.ua} %
\affiliation{%
National Science Center ``Kharkov Institute of Physics and
Technology'', 1, Akademicheskaya St., 61108 Kharkov, Ukraine }%

\begin{abstract}
A quantum-field approach for describing many-particle Fermi systems
at finite temperatures and with spontaneously broken symmetry has
been proposed. A generalized model of self-consistent field (SCF),
which allows one to describe the states eligible for this system
with various symmetries, is used as the initial approximation. A
perturbation theory has been developed, and a diagram technique for
temperature Green's functions (GFs) has been constructed. The
Dyson's equation for the self-energy and vertex parts has been
deduced.
\end{abstract}
\maketitle

A regular effective approach in studying the many-particle problem
at the microscopic level is the quantum-field approach borrowed by
the statistical physics of non-relativistic systems from quantum
field theory. This method was employed by Matsubara \cite{Matsubara}
to construct the thermodynamic perturbation theory and the diagram
technique for investigating many-particle systems at finite
temperatures. The Matsubara's approach was improved by Abrikosov,
Gor'kov, and Dzyaloshinskii \cite{AGD1,AGD2} and Fradkin
\cite{Fradkin}. The technique developed in these works is applicable
for systems, whose symmetry coincides with the symmetry of their
Hamiltonians. In order to describe the states of many-particle
systems with spontaneously broken symmetry (superfluidity,
superconductivity, ferromagnetism, crystals, etc.) and the phase
transitions between the states with various symmetries, the
quantum-field approach is to be modified. In spite of the progress
achieved in the development of the theory of many-particle systems
that are in the state with spontaneously broken symmetry, the
further development of the ways to describe such systems on the
basis of fundamental physical principles is a challenging direction
in the evolution of quantum statistical physics.

The approach proposed in this work is based on a consistent account,
already in the zero-order approximation, of the essential features
of the system state under consideration; first of all, its symmetry.
It can be made consistently only in the case where the interaction
between particles is made allowance for, at least to some extent,
not later than in the main approximation. A requirement for the
initial approximation to be simple dictates the necessity to choose
the SCF model as the main approximation. The relative simplicity of
this model is defined by the circumstance that it preserves, as well
as the model of independent particles, the one-particle (to say more
accurate, quasi-one-particle) description of the system as a gas
composed of quasiparticles which are characterized by individual
wave functions. An important element of the theory is making use of
the Bogolyubov concept of quasiaverages
\cite{Bogolyubov1,Bogolyubov2} which has been formulated, taking
into account the self-consistent approach, in work
\cite{Poluektov1}.

The SCF model is known to be widely used for calculating atomic
spectra \cite{Hartree}, the structure of atomic nuclei \cite{BBIG},
and the properties of molecules and solids \cite{Slater}; therefore,
one may hope that the properties of many-particle systems can be
calculated well enough by this method, with the quality of such
calculations growing as computer facilities are being developed. The
main attention in this work is given to the development, on the
basis of the generalized SCF model, of the quantum-field approach to
describe many-particle Fermi systems which are in states with
spontaneously broken symmetry (including spatially nonhomogeneous
ones) and to the construction of a perturbation theory and a diagram
technique. It has been shown that the thermodynamic functions found
from a microscopic consideration in the framework of the SCF model
obey the correct thermodynamic relationships, and, therefore, the
SCF model, which is fundamental for the theory proposed, is not
self-contradictory. It should be emphasized that the approach, which
is proposed for the consideration of many-particle systems, is based
only upon the general principles of non-relativistic quantum theory
and statistical mechanics and does not require additional
hypotheses.

{\bf 1.} The motion of a single particle in the external field
$U_0({\bf r})$ is described by the Schr\"{o}dinger equation
\begin{equation} \label{EQ01}
\begin{array}{l}
\displaystyle{%
  \int\! dq' H_0(q,q')\varphi_j(q')=\varepsilon_j^{(0)}\varphi_j(q)\,, %
}%
\end{array}
\end{equation}
where the notation $q=\{{\bf r}, \sigma\}$ stands for the set of
spatial, ${\bf r}$, and discrete spin, $\sigma$, coordinates; the
integration includes the summation over spin indices; the subscript
$j$ comprises the full set of quantum numbers, including the spin
projection, which characterizes the  state of an individual
particle; $\varphi_j(q)$ is the wave function of the particle; and
$\varepsilon_j^{(0)}$ is its energy. The kernel in Eq.\,(\ref{EQ01})
looks like
\begin{equation} \nonumber
\begin{array}{c}
\displaystyle{%
  H_0(q,q')\varphi_j(q')=-\frac{\hbar^2}{2m}\Delta\,\delta(q-q')+U_0({\bf r})\delta(q-q')\,,  %
}%
\end{array}
\end{equation}
where $\delta(q-q')=\delta({\bf r}-{\bf r}')\,\delta_{\sigma\sigma'}$, %
$m$ is the particle's mass, and $\Delta$ is the Laplacian. We will
use the methods of secondary quantization, having defined the
operators of creation, $a_j^+$, and annihilation, $a_j$, of real
Fermi particles. Let us also define the field operators
\begin{equation} \label{EQ02}
\begin{array}{l}
\displaystyle{%
  \Psi(q)=\sum_j \varphi_j(q)a_j, \quad  \Psi^+(q)=\sum_j \varphi_j^*(q)a_j^+\,. %
}%
\end{array}
\end{equation}
The particle operators and field operators (\ref{EQ02}) obey the
notorious anticommutation relations \cite{Bogolyubov2}. The
Hamiltonian of a many-particle Fermi system, expressed in terms of
the field operators, looks like
\begin{equation} \label{EQ03}
\begin{array}{ll}
\displaystyle{%
  H=\int\! dq\,dq'\,\Psi^+(q)H(q,q')\Psi(q')+  %
} \vspace{2mm}\\ %
\displaystyle{%
  \hspace{5mm}+\frac{1}{2}\int\! dq\,dq'\,\Psi^+(q)\Psi^+(q')U({\bf r},{\bf r}')\Psi(q')\Psi(q)\,,  %
}
\end{array}
\end{equation}
where
\begin{equation} \nonumber
\begin{array}{ll}
\displaystyle{%
  H(q,q')=H_0(q,q')-\mu\,\delta(q-q')\,,  %
} \vspace{2mm}\\ %
\end{array}
\end{equation}
and $U({\bf r},{\bf r}')$ is the potential of two-particle
interaction independent of spin variables. While studying
many-particle systems with broken symmetry, it is convenient to
suppose the considered system to be in contact with a thermostat and
to have an opportunity to exchange both energy and particles with
it; i.e. the total energy and the total number of particles in the
system are not regarded fixed. The thermostat is characterized by
two parameters: the temperature $T$ and the chemical potential
$\mu$. In the state of thermodynamic equilibrium, the same
parameters characterize the system of particles as well. Thus, we
use the grand canonical ensemble and deal with the Hamiltonian that
includes the term with the chemical potential $-\mu N$, where $N$ is
the operator of the number of particles.

{\bf 2.} To pass to the SCF model, let us present the initial
Hamiltonian (\ref{EQ03}) as a sum of two terms:
\begin{equation} \label{EQ04}
\begin{array}{l}
\displaystyle{%
  H=H_0+H_C\,, %
}
\end{array}
\end{equation}
where the first term is the Hamiltonian of the SCF model which is
quadratic in the field operators,
\begin{equation} \label{EQ05}
\begin{array}{ll}
\displaystyle{%
  H_0=\int\! dq\,dq'\biggl\{ \Psi^+(q)[H(q,q')+W(q,q')\Psi(q')]+
} \vspace{2mm}\\ %
\displaystyle{%
  \hspace{1mm} +\frac{1}{2}\Psi^+(q)\Delta(q,q')\Psi^+(q')\!+\!\frac{1}{2}\Psi(q')\Delta^*(q,q')\Psi(q)\!\biggr\}\!+E_0', %
}
\end{array}
\end{equation}
and the second one is the correlation Hamiltonian
\begin{equation} \label{EQ06}
\begin{array}{ll}
\displaystyle{%
  H_C=\frac{1}{2}\int\! dq\,dq'\biggl\{ \Psi^+(q)\Psi^+(q')U({\bf r},{\bf r}')\Psi(q')\Psi(q)- %
} \vspace{2mm}\\ %
\displaystyle{%
  \hspace{1mm} -2\Psi^+(q)W(q,q')\Psi(q')- %
} \vspace{2mm}\\ %
\displaystyle{%
  \hspace{1mm} -\Psi^+(q)\Delta(q,q')\Psi^+(q')-\Psi(q')\Delta^*(q,q')\Psi(q)   \!\biggr\}\!-E_0', %
}
\end{array}
\end{equation}
which takes those interactions into consideration that were not
accounted for in the SCF approximation. Formulae (\ref{EQ05}) and
(\ref{EQ06}) include self-consistent, still unknown potentials
$W(q,q')$ and $\Delta(q,q')$ which satisfy, owing to the
self-consistency of the Hamiltonian, the conditions
\begin{equation} \label{EQ07}
\begin{array}{ll}
\displaystyle{%
  W(q,q')=W^*(q',q), \quad \Delta(q,q')=-\Delta(q',q), %
}
\end{array}
\end{equation}
as well as the non-operator term $E_0'$, the choice of which is
essential for the thermodynamics of the considered model to be
constructed correctly. Therefore, in the SCF model, Hamiltonian
(\ref{EQ03}) is replaced by the simpler model Hamiltonian $H_0$
(\ref{EQ05}). The latter contains potentials which will be
determined from the condition of the best approximation of the exact
Hamiltonian $H$ by the model Hamiltonian $H_0$. An important
qualitative difference between these two Hamiltonians consists in
that the initial Hamiltonian does not depend on the system state,
whereas the self-consistent Hamiltonian $H_0$, as will be shown,
depends on the system state and the thermodynamic variables through
the self-consistent potentials $W(q,q')$ and $\Delta(q,q')$. It is
this property of the self-consistent Hamiltonian that allows one to
describe the states with broken symmetry. When constructing the
perturbation theory for many-particle systems with broken symmetry,
it is natural that the self-consistent Hamiltonian $H_0$ should be
chosen as the basic one, and the correlation Hamiltonian $H_C$ as a
perturbation.

With the help of Bogolyubov's canonical $u$-$v$-transformations
\begin{equation} \label{EQ08}
\begin{array}{ll}
\displaystyle{%
  \Psi(q)= \sum_i \left[ u_i(q)\gamma_i + v_i^*(q)\gamma_i^+ \right]\,, %
} \vspace{1mm}\\ %
\displaystyle{%
  \Psi^+(q)= \sum_i \left[ v_i(q)\gamma_i + u_i^*(q)\gamma_i^+ \right]\,, %
}
\end{array}
\end{equation}
we can diagonalize the self-consistent Hamiltonian (\ref{EQ05}):
\begin{equation} \label{EQ09}
\begin{array}{ll}
\displaystyle{%
  H_0=E_0 + \sum_i \varepsilon_i\, \gamma_i^+\gamma_i\,. %
}
\end{array}
\end{equation}
Here, $E_0$ is the non-operator part of the Hamiltonian;
$\varepsilon_i$ is the energy of elementary excitations,
quasiparticles, reckoned from the chemical potential level; $i$ is
the full set of quantum numbers, including the spin projection,
which characterizes the quasiparticle state; and the operators
$\gamma_i^+$ and $\gamma_i$ describe the processes of creation and
annihilation of quasiparticles. The quasiparticle description is
widely used in condensed mater physics. In the SCF model, the notion
of quasiparticles, which possess the infinite lifetime in this
approximation, arises naturally as a result of the reduction of
Hamiltonian (\ref{EQ05}) to the diagonal form (\ref{EQ09}). The
relative simplicity of such a model lies in that it retains the
one-particle (to say more precisely, one-quasiparticle) description
of the system. The set of factors $\{ u_i(q), v_i(q)\}$ is the
two-component wave function of a quasiparticle. For the transition
from the initial self-consistent Hamiltonian (\ref{EQ05}) to the
diagonalized one (\ref{EQ09}) to be feasible, the factors in
canonical transformations (\ref{EQ08}) must satisfy the
Bogolyubov-de Gennes system of equations \cite{Bogolyubov3,Gennes}
which looks, in the most general case, like
\begin{equation} \label{EQ10}
\begin{array}{ll}
\displaystyle{%
  \int\!dq'\left[ \Omega(q,q')u_i(q')+\Delta(q,q')v_i(q') \right]=\varepsilon_i\,u_i(q)\,,  %
}
\end{array}
\end{equation}
\vspace{-5mm}%
\begin{equation} \label{EQ11}
\begin{array}{ll}
\displaystyle{%
  \int\!dq'\left[ \Omega^*(q,q')v_i(q')+\Delta^*(q,q')u_i(q') \right]=-\varepsilon_i\,v_i(q)\,,  %
}
\end{array}
\end{equation}
where $\Omega(q,q')=H(q,q')+W(q,q')$. The requirement that for
transformations (\ref{EQ08}) to be canonical results in the
normalization and completeness of the solutions of the
self-consistent equations (\ref{EQ10}) and (\ref{EQ11})
\cite{Poluektov2}. The average values of operators in the SCF model
are expressed in terms of the normal, $\rho$, and abnormal, $\tau$,
one-particle density matrices:
\begin{equation} \label{EQ12}
\begin{array}{ll}
\displaystyle{%
  \rho(q,q')=\langle\Psi^+(q')\Psi(q)\rangle_0=   %
}\vspace{2mm}\\ %
\displaystyle{ \hspace{2mm}%
  =\sum_i [u_i(q)u_i^*(q')\,f_i + v_i^*(q)v_i(q')(1-f_i)]\,, %
}
\end{array}
\end{equation}
\vspace{-5mm}%
\begin{equation} \label{EQ13}
\begin{array}{ll}
\displaystyle{%
  \tau(q,q')=\langle\Psi(q')\Psi(q)\rangle_0=   %
}\vspace{2mm}\\ %
\displaystyle{ \hspace{2mm}%
  =\sum_i [u_i(q)v_i^*(q')\,f_i + v_i^*(q)u_i(q')(1-f_i)]\,, %
}
\end{array}
\end{equation}
where the quasiparticle distribution function $f_i$ has the same
form as in the model of ideal gas,
\begin{equation} \label{EQ14}
\begin{array}{ll}
\displaystyle{%
  f_i=f(\varepsilon_i)=(\exp\beta\varepsilon_i+1)^{-1}\,,  %
}
\end{array}
\end{equation}
and $\beta=1/T$ is the inverse temperature. Since the quasiparticle
energy is the functional of $f_i$, formula (\ref{EQ14}) turns out a
complicated nonlinear equation for the distribution function, being
similar to that which takes place in the Landau phenomenological
theory of a Fermi liquid \cite{Landau}. In formulae (\ref{EQ12}) and
(\ref{EQ13}), the averaging is carried on with the statistical
operator
\begin{equation} \label{EQ15}
\begin{array}{ll}
\displaystyle{%
  \rho_0=\exp\beta (\Omega_0-H_0)\,,  %
}
\end{array}
\end{equation}
where the normalizing constant $\Omega_0=-T\ln\textrm{Sp}e^{-\beta H_0}$ 
is determined by the condition $\textrm{Sp}\,\rho_0=1$ and has the
meaning of the thermodynamic potential of the system in the SCF
model.

In order that the system of equations (\ref{EQ10}) and (\ref{EQ11})
be completely defined, one has to express the self-consistent
potentials (\ref{EQ07}) in terms of the functions $u(q)$ and $v(q)$.
Using the variational principle \cite{Poluektov2}, we find the
connection between the self-consistent potentials with the
one-particle density matrices (\ref{EQ12}) and (\ref{EQ13}):
\begin{equation} \label{EQ16}
\begin{array}{ll}
\displaystyle{%
  W(q,q')=-U({\bf r},{\bf r}')\,\rho(q,q')\, +  %
}\vspace{2mm}\\ %
\displaystyle{ \hspace{2mm}%
  +\,\delta(q-q')\int\!\!dq''\,U({\bf r},{\bf r}'')\rho(q'',q'')\,,  %
}
\end{array}
\end{equation}
\vspace{-5mm}%
\begin{equation} \label{EQ17}
\begin{array}{ll}
\displaystyle{%
  \Delta(q,q')=U({\bf r},{\bf r}')\tau(q,q')\,.  %
}
\end{array}
\end{equation}
Substituting Eqs.\,(\ref{EQ16}) and (\ref{EQ17}) into
Eqs.\,(\ref{EQ10}) and (\ref{EQ11}) gives rise to a closed system of
nonlinear integro-differential equations for the wave functions
$u(q)$ and $v(q)$:
\begin{equation} \label{EQ18}
\begin{array}{ll}
\displaystyle{%
  -\frac{\hbar^2}{2m}\Delta u_i(q)+\!\biggl[U({\bf r})-\mu-\varepsilon_i\, + %
}\vspace{2mm}\\ %
\displaystyle{ \hspace{2mm}%
  +\int\!\!dq'\,U({\bf r},{\bf r}')\rho(q',q')\biggr]u_i(q)-  %
}\vspace{2mm}\\ %
\displaystyle{ \hspace{2mm}%
  -\int\!\!dq'\,U({\bf r},{\bf r}')[\rho(q,q')u_i(q')-\tau(q,q')v_i(q')]=0\,,  %
}
\end{array}
\end{equation}
\vspace{-5mm}%
\begin{equation} \label{EQ19}
\begin{array}{ll}
\displaystyle{%
  -\frac{\hbar^2}{2m}\Delta v_i(q)+\!\biggl[U({\bf r})-\mu+\varepsilon_i\, + %
}\vspace{2mm}\\ %
\displaystyle{ \hspace{2mm}%
  +\int\!\!dq'\,U({\bf r},{\bf r}')\rho(q',q')\biggr]v_i(q)-  %
}\vspace{2mm}\\ %
\displaystyle{ \hspace{2mm}%
  -\int\!\!dq'\,U({\bf r},{\bf r}')[\rho^*(q,q')v_i(q')-\tau^*(q,q')u_i(q')]=0\,.  %
}
\end{array}
\end{equation}
The chemical potential $\mu$ is related to the average number of
particles $N$ as
\begin{equation} \label{EQ20}
\begin{array}{ll}
\displaystyle{%
  N=\int\!d^{3}r\,n({\bf r}),\,\,\, n({\bf r})=\sum_\sigma\rho(q,q)\,,  %
}
\end{array}
\end{equation}
where $n({\bf r})$ is the particle number density. The system of
equations (\ref{EQ18}) and (\ref{EQ19}), together with relations
(\ref{EQ14}) and (\ref{EQ20}), describes fermion systems at finite
temperatures in the SCF approximation, being also valid when the
symmetry of the system states is spontaneously broken. In
particular, these equations are applicable for describing both
magnetic properties and superfluid (superconducting for charged
particles) states. For spatially nonhomogeneous states, where the
characteristic variation length of the functions $u(q)$ and $v(q)$
is much longer than the effective range of the interparticle
potential, Eqs.\,(\ref{EQ18}) and (\ref{EQ19}) can be presented in
the differential form. The constant $E_0'$ in Eq.\,(\ref{EQ05}) is
defined by the formula \cite{Poluektov2}
\begin{equation} \label{EQ21}
\begin{array}{ll}
\displaystyle{%
  E_0'\!=\!-\frac{1}{2}\int\!dq\,dq'\,U({\bf r},{\bf r}') \langle\Psi^+(q)\Psi^+(q')\Psi(q')\Psi(q)\rangle_0= %
}\vspace{2mm}\\ %
\displaystyle{ \hspace{2mm}%
  =\!\frac{1}{2}\int\!dq\,dq'\,U({\bf r},{\bf r}')\times  %
}\vspace{2mm}\\ %
\displaystyle{ \hspace{2mm}%
  \times\Big[ |\rho(q,q')|^2-\rho(q,q)\rho(q',q')-|\tau(q,q')|^2 \Big]\,.  %
}
\end{array}
\end{equation}
In the SCF approximation, the average value of the exact Hamiltonian
is equal to the average value of the self-consistent Hamiltonian: %
$\langle H\rangle_0=\langle H_0\rangle_0$.

In many cases, there is no necessity in knowing the quasiparticle
wave functions in order to find the equilibrium characteristics of
the researched system; the one-particle density matrices will do.
From Eqs.\,(\ref{EQ18}) and (\ref{EQ19}), as well as formulae
(\ref{EQ12}) and (\ref{EQ13}), the system of equations for the
one-particle density matrices reads:
\begin{equation} \label{EQ22}
\begin{array}{ll}
\displaystyle{%
  \frac{\hbar^2}{2m}[\Delta\rho(q,q')-\Delta'\rho(q,q')]-[U({\bf r})-U({\bf r}')]\rho(q,q')+ %
}\vspace{2mm}\\ %
\displaystyle{ \hspace{2mm}%
  +\int\!\!dq''\Big[ U({\bf r},{\bf r}'')-U({\bf r}',{\bf r}'')\Big]\Big[ \rho(q,q'')\rho(q'',q')-   %
}\vspace{2mm}\\ %
\displaystyle{ \hspace{2mm}%
  -\rho(q'',q'')\rho(q,q')+ \tau(q,q'')\tau^*(q'',q')\Big] = 0\,,  %
}
\end{array}
\end{equation}
\vspace{-5mm}%
\begin{equation} \label{EQ23}
\begin{array}{ll}
\displaystyle{%
  \frac{\hbar^2}{2m}[\Delta\tau(q,q')+\Delta'\tau(q,q')]- %
}\vspace{2mm}\\ %
\displaystyle{ \hspace{2mm}%
  -[U({\bf r})+U({\bf r}')+U({\bf r},{\bf r}')-2\mu]\,\tau(q,q')+  %
}\vspace{2mm}\\ %
\displaystyle{ \hspace{2mm}%
  +\int\!\!dq''\Big[ U({\bf r},{\bf r}'')+U({\bf r}',{\bf r}'')\Big]\Big[ \rho(q,q'')\tau(q'',q')- %
}\vspace{2mm}\\ %
\displaystyle{ \hspace{2mm}%
  -\rho(q'',q'')\tau(q,q')+\tau(q,q'')\rho(q',q'')\Big]=0\,.  %
}
\end{array}
\end{equation}
The solutions of this system and, hence, the corresponding
self-consistent fields (\ref{EQ16}) and (\ref{EQ17}) can possess
different symmetries, including that which is lower than the
symmetry of the initial Hamiltonian (\ref{EQ03}), and thus can
describe states with a spontaneously broken symmetry. In particular,
the system of equations (\ref{EQ22}) and (\ref{EQ23}) has both
normal solutions, for which $\tau(q,q')=0$ and $\rho(q,q')\neq 0$,
and ``superfluid'' ones, for which both $\tau(q,q')$ and
$\rho(q,q')$ do not vanish. The anomalous density matrix
$\tau(q,q')$ or the self-consistent potential $\Delta(q,q')$ can be
considered as the microscopic order parameters of the superfluid
state. If the dependencies of the one-particle density matrices on
the spin variables are such that
$\rho(q,q')\propto\delta_{\sigma\sigma'}$ and %
$\tau(q,q')\propto\sigma^{(y)}_{\sigma\sigma'}$ %
($\hat{\sigma}^{(y)}$ is the Pauli spin matrix), the system is
invariant with respect to spin rotations. Otherwise, there is a
magnetic ordering in the many-particle system. If the thermodynamic
parameters are fixed, only that state among the possible ones of the
system will be realized really, whose thermodynamic potential is
minimal.

{\bf 3.} A distinctive feature of the SCF model, which should be
taken into account when deriving thermodynamic relations from
Hamiltonian (\ref{EQ05}), is that this Hamiltonian contains
self-consistent potentials and an operator-free term which depend on
the temperature and the chemical potential. Only the correct choice
of the self-consistent potentials $W$ and $\Delta$ and the quantity
$E_0'$ ensures that the thermodynamic relations would be satisfied.
Using the definitions of thermodynamic potential (\ref{EQ15}) and
entropy $S_0=-\textrm{Sp}(\rho_0\ln\rho_0)$, one can demonstrate
\cite{Poluektov2,Poluektov3} that the thermodynamic relation
$\Omega_0=E-TS_0-\mu N$, where $E$ is the total energy of the
system, is satisfied, and the variation of the thermodynamic
potential is equal to the averaged variation of $H_0$:
\begin{equation} \label{EQ24}
\begin{array}{ll}
\displaystyle{%
  \delta\Omega_0=\langle\delta H_0\rangle_0\,.  %
}
\end{array}
\end{equation}
By varying the self-consistent Hamiltonian which is expressed in
terms of one-particle density matrices and taking Eq.\,(\ref{EQ24})
into account, we obtain
\begin{equation} \label{EQ25}
\begin{array}{ll}
\displaystyle{%
 \frac{\delta\Omega_0}{\delta\rho(q,q')}\!=\!\bigg\langle\!\frac{\delta H_0}{\delta \rho(q,q')}\!\bigg\rangle_0  %
 \!=\!\frac{\delta\Omega_0}{\delta\tau^*(q,q')}\!=\!\bigg\langle\!\frac{\delta H_0}{\delta\tau^*(q,q')}\!\bigg\rangle_0\!= 0. %
}
\end{array}
\end{equation}

The relations (\ref{EQ16}) and (\ref{EQ17}) between the potentials
$W$ and $\Delta$ and the one-particle density matrices, which have
been established with the help of the variational principle, make
the thermodynamic potential extremal, as is seen from
Eq.\,(\ref{EQ25}), with respect to its variation over the
one-particle density matrices $\rho$ and $\tau$. Due to
Eq.\,(\ref{EQ25}), in the case where the volume of the system is
fixed, the ordinary thermodynamic relation
\begin{equation} \label{EQ26}
\begin{array}{ll}
\displaystyle{%
 d\Omega_0=-S_0 dT - N d\mu %
}
\end{array}
\end{equation}
is satisfied. The total energy can be found either by averaging the
energy operator directly or with the help of the thermodynamic
relation in terms of the thermodynamic potential
\begin{equation} \label{EQ27}
\begin{array}{ll}
\displaystyle{%
  E=\Omega_0 -\mu\frac{\partial\Omega_0}{\partial\mu}-T\frac{\partial\Omega_0}{\partial T}\,.  %
}
\end{array}
\end{equation}
According to Eqs.\,(\ref{EQ26}) and (\ref{EQ27}), the fact that the
self-consistent Hamiltonian involves the potentials depending on
thermodynamic variables does not violate thermodynamic relations, as
might have appeared \cite{Kirzhnits}, so that the SCF approximation
in statistics is intrinsically non-contradictory.

The total energy of the system of Fermi particles in the SCF model
is a sum of several contributions (the kinetic energy $K$, energy of
particles in an external field $U_E$, energy of direct
particle-particle interaction $U_D$, energy of exchange interaction
$U_{\textrm{ex}}$, and energy of condensation into a superfluid
state $U_C$) and can be written down in the form
\begin{equation} \label{EQ28}
\begin{array}{ll}
\displaystyle{%
  E\!=\!\sum_i \varepsilon_i f_i\!-\!\sum_i\varepsilon_i\!\int\!\!dq|v_i(q)|^2\!+\!\mu N\!-\!(U_D\!+\!U_{\textrm{ex}}\!+\!U_C).  %
}
\end{array}
\end{equation}
One can readily see that the total energy is not the sum of energies
of individual quasiparticles. Carrying out averaging in
Eq.\,(\ref{EQ09}) and taking into account Eq.\,(\ref{EQ28}), we
obtain the constant $E_0$ in the diagonalized Hamiltonian
(\ref{EQ09}):
\begin{equation} \label{EQ29}
\begin{array}{ll}
\displaystyle{%
  E=-(U_D\!+\!U_{\textrm{ex}}\!+\!U_C)-\sum_i\varepsilon_i\!\int\!\!dq|v_i(q)|^2\,.  %
}
\end{array}
\end{equation}
Now, the ultimate form of the thermodynamic potential in the SCF
approximation can be found easily as
\begin{equation} \label{EQ30}
\begin{array}{ll}
\displaystyle{%
  \Omega_0=-(U_D\!+\!U_{\textrm{ex}}\!+\!U_C)-\sum_i\varepsilon_i\!\int\!\!dq|v_i(q)|^2 -   %
}\vspace{2mm}\\ %
\displaystyle{ \hspace{2mm}%
  - T\sum_i\ln\big(1+e^{-\beta\varepsilon_i}\big)\,.  %
}
\end{array}
\end{equation}
In formulae (\ref{EQ28})\! -- \!(\ref{EQ30}),
\begin{equation} \nonumber
\begin{array}{ll}
\displaystyle{%
  U_D= \frac{1}{2}\int\!\! d^3r\,d^3r'\,U({\bf r},{\bf r}')\,n({\bf r})\,n({\bf r}')\,, %
}\vspace{3mm}\\ %
\displaystyle{ %
  U_{\textrm{ex}}= -\frac{1}{2}\int\!\! dq\,dq'\,U({\bf r},{\bf r}')\,|\rho(q,q')|^2 \,, %
}\vspace{3mm}\\ %
\displaystyle{ %
  U_{C}= \frac{1}{2}\int\!\! dq\,dq'\,U({\bf r},{\bf r}')\,|\tau(q,q')|^2 \,.  %
}
\end{array}
\end{equation}

The total number of quasiparticles is defined by the relation
\begin{equation} \label{EQ31}
\begin{array}{ll}
\displaystyle{%
  N_q= \sum_i f_i\,,  %
}
\end{array}
\end{equation}
while the total number of particles looks like
\begin{equation} \label{EQ32}
\begin{array}{ll}
\displaystyle{%
  N=N_q + \sum_i \textrm{th}\frac{\beta\varepsilon_i}{2}\int\!\!dq|v_i(q)|^2\,.  %
}
\end{array}
\end{equation}
Hence, in the normal (non-superfluid) state the number of particles
is equal to the number of quasiparticles, similarly to what takes
place in the theory of a normal Landau Fermi liquid \cite{Landau}.
In the superfluid state, as follows from Eq.\,(\ref{EQ32}), the
number of quasiparticles is always less than the number of
particles. It can be interpreted in a way that a certain number of
particles in the superfluid phase forms the condensate of Cooper
pairs and does not contribute to the formation of quasiparticle
excitations.

The relations of the SCF theory quoted above can be presented
\cite{Poluektov4} in the form of the relations of the Fermi liquid
theory \cite{Landau}. The equations of the generalized SCF model
formulated here lead to the results of the BCS theory of
superconductivity \cite{BCS}, as well as the relations of the theory
of a superfluid Fermi liquid which has been constructed in works
\cite{KPY,AKPY}. The equations of the generalized SCF model are
capable to describe magnetic properties of collective electrons and
bring about the relations of the Stoner theory \cite{Stoner}. Note
that the proposed theory makes it possible to write down the known
Stoner criterion for ferromagnetism in another form,
$I\rho(\varepsilon_0)>1$, where $I$ is the parameter of exchange
interaction and $\rho(\varepsilon_0)$ is the density of states at
the Fermi level. In the case of point-like interaction between
particles, $U({\bf r},{\bf r}')=U_0\delta({\bf r}-{\bf r}')$, this
criterion looks like
\begin{equation} \label{EQ33}
\begin{array}{ll}
\displaystyle{%
  \frac{a}{l}> \xi\,,  %
}
\end{array}
\end{equation}
where $l=(V/N)^{1/3}$ is the average distance between particles,
$a=U_0m/(4\pi\hbar^2)$ is the scattering length, and
$\xi=\frac{1}{2}(\frac{\pi}{3})^{1/3}\approx 0.5$ is a numerical
parameter. The approach proposed here enables the competition of
superconductivity and magnetism, as well as spatially modulated
states, to be studied.

{\bf 4.} While studying theoretically the systems with broken
symmetry, one can not take advantage of the conventional definition
of the average for calculating observable characteristics, because
the symmetry of a state with broken symmetry is lower than that of
its Hamiltonian. At the same time, when calculating the averages
according to the routines of statistical mechanics, the symmetry of
the averages coincides with that of the Hamiltonian. Such a
contradiction does not arise in the SCF model, because the system of
self-consistent equations has solutions with symmetry lower than
that of the initial Hamiltonian. To overcome those difficulties,
Bogolyubov \cite{Bogolyubov1} introduced the concept of
quasiaverages into statistical mechanics. According to this concept,
the averages in states with broken symmetry should be calculated not
using Hamiltonian (\ref{EQ03}) but a Hamiltonian, which differs from
(\ref{EQ03}) by terms that violate its symmetry in an appropriate
way. In the framework of such an approach, however, some uncertainty
in the introduction of fields that violate symmetry remains. As the
choice of symmetry-violating fields does not depend on interparticle
interactions, it may happen that the interactions do not allow the
existence of states possessing the symmetry imposed by the field
introduced. A way to determine quasiaverages using the
self-consistent Hamiltonian as an additive that violates the
symmetry was proposed in work \cite{Poluektov1}. In this case, the
system can possess only the symmetry that is allowed by
interparticle interactions.

Though the symmetry of each of the Hamiltonians $H_0$ and $H_C$,
which depend on the system state, can be lower than the symmetry of
the initial Hamiltonian, the symmetry of $H$ does not depend,
naturally, on the way how it was split and, thus, remains invariant.
Therefore, in order to describe the systems with broken symmetry,
let us introduce a more general Hamiltonian
\begin{equation} \label{EQ34}
\begin{array}{ll}
\displaystyle{%
  H_g = H_0 + g H_C  %
}
\end{array}
\end{equation}
which depends on a real positive parameter $g$. It is obvious that,
at $g=1$, this Hamiltonian coincides with the initial one
(\ref{EQ03}) and, at $g=0$, with the self-consistent Hamiltonian
(\ref{EQ05}). The variation of this parameter from zero to unity
means the inclusion of the correlation interaction. If $g$ is very
close to unity, Hamiltonian (\ref{EQ34}) practically coincides with
the initial one (\ref{EQ03}). The major difference, however,
consists in that the symmetry of Hamiltonian (\ref{EQ34}) coincides
at that with the symmetry of the self-consistent Hamiltonian and can
be lower than the symmetry of the initial Hamiltonian. Let us define
the statistical operator
\begin{equation} \label{EQ35}
\begin{array}{ll}
\displaystyle{%
  \rho_g = e^{\beta(\Omega_g-H_g)}, \quad  \Omega_g=-T\ln\textrm{Sp}e^{-\beta H_g}\,. %
}
\end{array}
\end{equation}
The quasiaverage of an arbitrary operator will be determined by the
relation
\begin{equation} \label{EQ36}
\begin{array}{ll}
\displaystyle{%
  \langle A\rangle = \lim_{g\rightarrow 1}\lim_{V\rightarrow\infty}\textrm{Sp}\,\rho_g A\,. %
}
\end{array}
\end{equation}
If the values of the thermodynamic variables $\mu$ and $T$ are
fixed, quasiaverages (\ref{EQ36}) can be not equal to conventional
averages and, therefore, cannot describe the states with broken
symmetry. From the mathematical point of view, a possible divergence
between averages and quasiaverages is known \cite{Bogolyubov1,AP} to
arise from the dependence of the result on the sequence of the
passages to the limit in Eq.\,(\ref{EQ36}). The passage to the limit
of the ``coupling constant'' $g$ must be carried out after the
thermodynamic passage to the limits $V\rightarrow\infty$ and
$N\rightarrow\infty$, provided $N/V=\textrm{const}$. If the symmetry
is not broken, quasiaverages (\ref{EQ36}) are identical to the
relevant conventional averages.

{\bf 5.} Correlation Hamiltonian (\ref{EQ06}) which we have selected
as a perturbation has a rather complicated structure. However, it
can be written down in a much more compact form, and the
perturbation theory will accept a simpler form if one uses the
concept of the normal product of operators. This concept plays an
essential role in quantum field theory \cite{BS}. In the
temperature-involved technique which was put forward in works
\cite{Matsubara,AGD1,AGD2,Fradkin}, the concept of the normal
product is not applied; therefore, the analogy with quantum field
theory is incomplete.

To define the normal product of Fermi operators, one has to pass
preliminarily to the particle-hole representation. Such a transition
at $T=0$ does not meet difficulties \cite{Kirzhnits}. At non-zero
temperatures, this procedure is not so obvious, because one-particle
states are not divided unambiguously into occupied and free ones; so
that every state may be either free or occupied with a certain
probability. However, the concept of the normal product can be
generalized for finite temperatures so that it will be independent
of the transition to the particle-hole representation.

For the further consideration, it is convenient to introduce the
notation for operators, using the ``isotopic'' index which acquires
two values, $1$ and $2$:
\begin{equation} \label{EQ37}
\begin{array}{ll}
\displaystyle{%
  a_{\alpha j}\!=\!\left\{\!
               \begin{array}{l}
                 a_j, \\
                 a_j^+,
               \end{array} \right. \,\,
  \gamma_{\alpha i}\!=\!\left\{\!
               \begin{array}{l}
                 \gamma_i, \\
                 \gamma_i^+,
               \end{array} \right. \,\,
  \Psi_\alpha(q)\!=\!\left\{\!
               \begin{array}{l}
                 \Psi(q), \\
                 \Psi^+(q),
               \end{array} \right. \,\,
               \begin{array}{l}
                 \alpha =1, \\
                 \alpha =2.
               \end{array}
}%
\end{array}
\end{equation}
We also use the notation $\bar{\alpha}$ which means
\begin{equation} \label{EQ38}
\begin{array}{ll}
\displaystyle{%
  \bar{\alpha}=\left\{
               \begin{array}{l}
                 1, \textrm{ if }\hspace{1mm} \alpha=2, \\
                 2, \textrm{ if }\hspace{1mm} \alpha=1.
               \end{array} \right.
}%
\end{array}
\end{equation}

The temperature normal product of two operators is defined by the
formula
\begin{equation} \label{EQ39}
\begin{array}{ll}
\displaystyle{%
  N(\Psi_1\Psi_2)=\Psi_1\Psi_2-\rho_{21}\,, %
}
\end{array}
\end{equation}
where $\rho_{21}=\langle\Psi_1\Psi_2\rangle_0$ is the one-particle
density matrix and $1=(q_1,i_1)$. It is evident that
\begin{equation} \label{EQ40}
\begin{array}{ll}
\displaystyle{%
  \langle N(\Psi_1\Psi_2)\rangle_0=0\,. %
}
\end{array}
\end{equation}

Let us give the general definition of the normal product of
operators which should be valid for both the Fermi and Bose
statistics. We introduce the notion of operator pairing which means
the average over a self-consistent state,
\begin{equation} \label{EQ41}
\begin{array}{ll}
\displaystyle{%
  \eta_1^a\eta_2^a = \langle\eta_1\eta_2\rangle_0\,,  %
}
\end{array}
\end{equation}
where $\eta_i$ is any of the operators $a_{\alpha i}, \Psi_{\alpha
i}$ or $\gamma_{\alpha i}$. The product of an arbitrary number of
operators containing the pairings is defined as
\begin{equation} \label{EQ42}
\begin{array}{ll}
\displaystyle{%
  \eta_1^a\eta_2\eta_3^a\eta_4\ldots\eta_k^b\ldots\eta_m^b\ldots\eta_{j-1}\eta_j =  %
}\vspace{2mm}\\ %
\displaystyle{ \hspace{2mm}%
  =a\langle\eta_1\eta_3\rangle_0 \langle\eta_k\eta_m\rangle_0\times %
}\vspace{2mm}\\ %
\displaystyle{ \hspace{2mm}%
  \times\, \eta_2\eta_4\ldots \eta_{k-1}\eta_{k+1}\ldots \eta_{m-1}\eta_{m+1}\ldots\eta_{j-1}\eta_j\,, %
}%
\end{array}
\end{equation}
where $a$ is a numerical factor which is equal to unity for Bose
operators and $(-1)^p$ for Fermi ones; and $p$ is the number of
permutations that are needed for putting the paired operators side
by side in the initial order. Taking into account the given
definition of pairings, the normal product of an arbitrary number of
operators is defined by the formula
\begin{equation} \label{EQ43}
\begin{array}{ll}
\displaystyle{%
  N(\eta_1\eta_2\ldots\eta_j)=\eta_1\eta_2\ldots\eta_j- %
}\vspace{2mm}\\ %
\displaystyle{ \hspace{2mm}%
  - \eta_1^a\eta_2^a\eta_3\ldots\eta_j - \eta_1^a\eta_2\eta_3^a\ldots\eta_j - %
}\vspace{2mm}\\ %
\displaystyle{ \hspace{2mm}%
  - (\textrm{all other products with one pairing}) +  %
}\vspace{2mm}\\ %
\displaystyle{ \hspace{2mm}%
  + \eta_1^a\eta_2^a\eta_3^b\eta_4^b\ldots\eta_j + \eta_1^a\eta_2^b\eta_3^a\eta_4^b\ldots\eta_j +  %
}\vspace{2mm}\\ %
\displaystyle{ \hspace{2mm}%
  + (\textrm{all other products with two pairings}) - \ldots \,\,\,.  %
}
\end{array}
\end{equation}
Thus, the temperature normal product of operators is defined as a
sum of operator products, which include all possible pairings
(including the term without pairings). The sigh plus is selected if
the number of pairings in the product is even, and the minus if odd.
The average of the $N$-product of any number of operators in the
Schr\"{o}dinger or interaction representation over a self-consistent
state is equal to zero,
\begin{equation} \label{EQ44}
\begin{array}{ll}
\displaystyle{%
  \langle N(\Psi_1\ldots\Psi_j\rangle_0 = 0\,,  %
}
\end{array}
\end{equation}
except the average of the $N$-product of a $c$-number which is
$N(c)=c$ by definition. We do not develop the general proof of
property (\ref{EQ44}) in detail, but it is easy to check in a
straightforward manner that it is fulfilled for the $N$-product of,
e.g., four operators:
\begin{equation} \label{EQ45}
\begin{array}{ll}
\displaystyle{%
  N(\Psi_1\Psi_2\Psi_3\Psi_4) = \Psi_1\Psi_2\Psi_3\Psi_4 - %
}\vspace{2mm}\\ %
\displaystyle{ \hspace{2mm}%
  - \Psi_1\Psi_2\rho_{43}-\Psi_3\Psi_4\rho_{21}-\Psi_2\Psi_3\rho_{41}-\Psi_1\Psi_4\rho_{32}+ %
}\vspace{2mm}\\ %
\displaystyle{ \hspace{2mm}%
  + \Psi_2\Psi_4\rho_{31} + \Psi_1\Psi_3\rho_{42}-\rho_{31}\rho_{42}+\rho_{21}\rho_{43}+\rho_{41}\rho_{32}. %
}
\end{array}
\end{equation}

An important property of the SCF model is that it allows a rather
complicated correlation Hamiltonian (\ref{EQ06}) to be represented
as the normal product of four field operators:
\begin{equation} \label{EQ46}
\begin{array}{ll}
\displaystyle{%
  H_C=\frac{1}{2}\int\!\!dqdq'\,U({\bf r},{\bf r}')\,N\big[\Psi^+(q)\Psi^+(q')\Psi(q')\Psi(q)\big]= %
}\vspace{2mm}\\ %
\displaystyle{ \hspace{2mm}%
  =\frac{1}{4}\sum_\alpha\int\!\!dqdq'\,U({\bf r},{\bf r}')\,N\big[\Psi_\alpha(q)\Psi_\alpha(q')\Psi_{\bar{\alpha}}(q')\Psi_{\bar{\alpha}}(q)\big].  %
}
\end{array}
\end{equation}
Similarly, this Hamiltonian can be expressed in terms of
quasiparticle operators:
\begin{equation} \label{EQ47}
\begin{array}{ll}
\displaystyle{%
  H_C=\frac{1}{2}\sum_{1234}U_{1234}\,N(\gamma_1\gamma_2\gamma_3\gamma_4)\,, %
}%
\end{array}
\end{equation}
where the matrix element of the interaction potential
\begin{equation} \label{EQ48}
\begin{array}{ll}
\displaystyle{%
  U_{1234}\!=\!\int\!\!dqdq'\,U({\bf r},{\bf r}')\,u_{i_1}^{2\alpha_1}(q)u_{i_2}^{2\alpha_2}(q)u_{i_3}^{1\alpha_3}(q)u_{i_4}^{1\alpha_4}(q)  %
}%
\end{array}
\end{equation}
is expressed through the coefficients of the Bogolyubov
transformation matrix
\begin{equation} \nonumber
\begin{array}{ll}
\displaystyle{%
  \hat{u}_i(q)= \left[
                  \begin{array}{cc}
                    u_i(q) & v_i^*(q) \\
                    v_i(q) & u_i^*(q) \\
                  \end{array}
                \right].
}%
\end{array}
\end{equation}
Thus, the Hamiltonian of the correlation interaction can be written
down in the form of a normal product not only for the normal systems
at zero temperature \cite{Kirzhnits} but also for the states with
spontaneously broken symmetry at finite temperatures.

{\bf 6.} An $L$-point temperature Green's function (GF) is defined as %
\begin{equation} \label{EQ49}
\begin{array}{ll}
\displaystyle{%
  G(1,2,\ldots L)= i^L \Big\langle T_\tau \hat{A}(1)\hat{A}(2)\ldots \hat{A}(L)\Big\rangle\,,  %
}%
\end{array}
\end{equation}
where the averaging should be understood as the quasiaveraging
(\ref{EQ36}), and each number stands for the whole set of variables.
The operators that are averaged in Eq.\,(\ref{EQ49}) are in the
Heisenberg-Matsubara representation:
\begin{equation} \label{EQ50}
\begin{array}{ll}
\displaystyle{%
  \hat{A}_\alpha(\tau)= e^{\tau H_g}A_\alpha e^{-\tau H_g}\,,  %
}%
\end{array}
\end{equation}
where $A_\alpha$ is the corresponding operator in the
Schr\"{o}dinger representation, $\tau$ the Matsubara ``time''
parameter $(0\le\tau\le\beta)$, and $T_\tau$ the operator of
chronological ordering \cite{Matsubara,AGD1}. If
$\hat{A}(1)\!=\!\hat{\Psi}(1)$, Eq.\,(\ref{EQ49}) defines an
$L$-point field GF, and if $\hat{A}(1)\!=\!\hat{\gamma}(1)$, an
$L$-point quasiparticle GF. When studying Fermi systems, only GFs
with even L's are considered. In particular, a two-point
(one-particle) GFs are defined by the formulae
\begin{equation} \label{EQ51}
\begin{array}{ll}
\displaystyle{%
  G^{\alpha\alpha'}(q\tau,q'\tau')=-\big\langle T_\tau\,\hat{\Psi}_\alpha(q\tau)\,\hat{\Psi}_{\alpha'}(q'\tau') \big\rangle\,,  %
}\vspace{3mm}\\ %
\displaystyle{ %
  \tilde{G}^{\alpha\alpha'}(i\tau,i'\tau')=-\big\langle T_\tau\,\hat{\gamma}_{\alpha i}(\tau)\,\hat{\gamma}_{\alpha' i'}(\tau') \big\rangle\,. %
}
\end{array}
\end{equation}
These functions are $4\times 4$ matrices in the spin and
``isotopic'' spaces. The components of GFs (\ref{EQ51}) that are
diagonal in isotopic indices are anomalous and distinct from zero
only in the superfluid state, while the non-diagonal ones are
non-zero both in the superfluid and normal states. To construct the
perturbation theory, the operators in the Matsubara interaction
representation are to be introduced:
\begin{equation} \label{EQ52}
\begin{array}{ll}
\displaystyle{%
  A_\alpha(\tau)= e^{\tau H_0}A_\alpha e^{-\tau H_0}\,.  %
}%
\end{array}
\end{equation}
Using these operators, we define the temperature GFs in the
framework of the SCF model by relations
\begin{equation} \label{EQ53}
\begin{array}{ll}
\displaystyle{%
  G^{(0)\alpha\alpha'}\!(q\tau,q'\tau')=-\big\langle T_\tau\,\Psi_\alpha(q\tau)\,\Psi_{\alpha'}(q'\tau') \big\rangle_0\,,  %
}\vspace{3mm}\\ %
\displaystyle{ %
  \tilde{G}^{(0)\alpha\alpha'}\!(i\tau,i'\tau')=-\big\langle T_\tau\,\gamma_{\alpha i}(\tau)\,\gamma_{\alpha' i'}(\tau') \big\rangle_0\,. %
}
\end{array}
\end{equation}
Here, the averaging is carried on with the statistical operator
(\ref{EQ15}) over the self-consistent state. Functions (\ref{EQ51})
and (\ref{EQ53}) depend only on the ``time'' difference $\tau-\tau'$
and satisfy the symmetry relations
\begin{equation} \label{EQ54}
\begin{array}{ll}
\displaystyle{%
  G^{\alpha\alpha'}\!(q,q';\tau-\tau')=-G^{\alpha'\alpha}(q',q;\tau'-\tau)=  %
}\vspace{3mm}\\ %
\displaystyle{ %
  = G^{\bar{\alpha}'\bar{\alpha}\,*}(q',q;\tau-\tau')=-G^{\bar{\alpha}\bar{\alpha}'*}(q,q';\tau'-\tau)\,. %
}
\end{array}
\end{equation}

For GFs (\ref{EQ53}) in the SCF model, the closed equations are
valid. It is most convenient to calculate the quasiparticle GF, for
which
\begin{equation} \label{EQ55}
\begin{array}{ll}
\displaystyle{%
  \frac{\partial \tilde{G}^{(0)\alpha\alpha'}\!(i\tau,i'\tau')}{\partial\tau}-\eta_\alpha\,\varepsilon_i\,\tilde{G}^{(0)\alpha\alpha'}\!(i\tau,i'\tau')=  %
}\vspace{3mm}\\ %
\displaystyle{ \hspace{2mm} %
  = -\delta_{ii'}\delta_{\alpha\bar{\alpha'}}\delta(\tau-\tau')\,,  %
}
\end{array}
\end{equation}
where $\eta_\alpha\!=\!(-1)^\alpha$. Expanding the GF in
Eq.\,(\ref{EQ55}) in a Fourier series, we find the Fourier component
\begin{equation} \label{EQ56}
\begin{array}{ll}
\displaystyle{%
  \tilde{G}_{ii'}^{(0)\alpha\alpha'}(\omega_n)= \frac{\delta_{\alpha\bar{\alpha'}}\delta_{ii'}}{i\omega_n+\eta_\alpha\varepsilon_i}\,, %
}
\end{array}
\end{equation}
where $\omega_n\!=\!\pi T(2n+1)$. Therefore, both in the normal and
superfluid states, only those terms are distinct from zero which are
non-diagonal in isotopic indices. Taking into account
Eq.\,(\ref{EQ08}), we find the GF in the field representation:
\begin{equation} \label{EQ57}
\begin{array}{ll}
\displaystyle{%
  G^{(0)\alpha\alpha'}\!(q,q';\omega_n)= \sum_{i,\alpha''}\frac{ u_i^{\alpha\alpha''}\!(q)\,u_i^{\alpha'\bar{\alpha''}}\!(q') }
  {i\omega_n+\eta_{\alpha''}\varepsilon_i}\,. %
}
\end{array}
\end{equation}
In this case, distinct from zero are the terms which are either
diagonal or non-diagonal in isotopic indices.

{\bf 7.} Let us construct the diagram technique to find the field
and quasiparticle GFs. Define the contraction of operators in the
interaction representation by the relation
\begin{equation} \label{EQ58}
\begin{array}{ll}
\displaystyle{%
  \Psi_\alpha^a(q,\tau)\Psi_{\alpha'}^a(q',\tau')=T_\tau(\Psi_\alpha(q,\tau)\Psi_{\alpha'}(q',\tau')) -  %
}\vspace{3mm}\\ %
\displaystyle{ \hspace{30mm} %
  -N(\Psi_\alpha(q,\tau)\Psi_{\alpha'}(q',\tau')).  %
}
\end{array}
\end{equation}
Contraction (\ref{EQ58}) is a $c$-number which coincides, with an
accuracy to a sign, with the GF
\begin{equation} \label{EQ59}
\begin{array}{ll}
\displaystyle{%
  \Psi_\alpha^a(q,\tau)\Psi_{\alpha'}^a(q',\tau')=-G^{(0)\alpha\alpha'}\!(q\tau,q'\tau').  %
}
\end{array}
\end{equation}

We note that product (\ref{EQ43}) and contraction (\ref{EQ58}) are
so defined that Wick's theorems remain valid in that form as they
were formulated in quantum field theory \cite{BS}; therefore, the
perturbation theory can be built in the standard way
\cite{AGD1,Kirzhnits,BS}. To present the diagram technique in a more
compact form, we designate variables, on which the GF depends, by a
single number, e.g., $1\!\equiv\!({\bf r}_1,\alpha_1,\sigma_1,\tau_1)$ %
and $\bar{1}\!\equiv\!({\bf r}_1,\bar{\alpha}_1,\sigma_1,\tau_1)$,
and consider the integration over the digital variable as the
integration over all continuous coordinates and the summation over
all discrete ones. In order to construct the perturbation theory, it
is necessary to express the temperature GFs in terms of operators in
the interaction representation and to pass from the averaging with
the statistical operator $\rho_g$ [see Eq.\,(\ref{EQ35})] to that
with the statistical operator $\rho_0$ [see Eq.\,(\ref{EQ15})].
Carrying out such a transition \cite{AGD1,Kirzhnits}, we obtain
\begin{equation} \label{EQ60}
\begin{array}{ll}
\displaystyle{%
  G^{\alpha\alpha'}\!(q\tau,q'\tau')=
  -\frac{\big\langle T_\tau\,\Psi_\alpha(q\tau)\Psi_{\alpha'}(q'\tau')\,\sigma(\beta) \big\rangle_0}
  {\langle \sigma(\beta)\rangle_0} %
}
\end{array}
\end{equation}
for the one-particle GF, where the temperature scattering matrix
looks like
\begin{equation} \label{EQ61}
\begin{array}{ll}
\displaystyle{%
  \sigma(\beta)=T_\tau\exp\!{\bigg[ -\int_0^\beta\!\! H_C(\tau')\,d\tau'\bigg]}\,, %
}
\end{array}
\end{equation}
where $H_C(\tau)\!=\!e^{\tau H_0}H_C\,e^{-\tau H_0}$ %
is the perturbation Hamiltonian in the interaction representation.
Since the perturbation Hamiltonians (\ref{EQ46}) and (\ref{EQ47})
are presented in the normally ordered form, the averaging over the
self-consistent state is to be applied to the $T$-products of
operators that are combined in fours under the symbol of the
$N$-product. This reduces the number of pairings substantially and,
accordingly, simplifies the diagram technique. In principle, one
could construct the theory without resorting to the normal ordering
of operators. In this case, the theory would contain a large number
of\, ``superfluous'' diagrams which would not contribute to the
final result, being reduced in every order of perturbation theory.
Using the $N$-ordered form of the perturbation Hamiltonian allows
the appearance of such diagrams to be excluded. In this case,
similarly to the standard technique \cite{AGD1,Kirzhnits,BKY}, the
theorem on connectivity is valid, so that the denominator in
Eq.\,(\ref{EQ60}) should not be taken into account, and only
connected diagrams should be allowed for in the numerator.

We note that the total thermodynamic potential of the system is
expressed in terms of the temperature scattering matrix averaged
over the self-consistent state. This averaged quantity can be
presented in the form \cite{AGD1,BKY}
\begin{equation} \nonumber
\begin{array}{ll}
\displaystyle{%
   \langle \sigma(\beta)\rangle_0 = \exp{ \!\left[ \sum_{n=0}^{\infty} \langle \sigma_n(\beta)\rangle_{0\,\textrm{conn}} \right] }\,,   %
}
\end{array}
\end{equation}
so that the total thermodynamic potential is defined by the formula
\begin{equation} \label{EQ62}
\begin{array}{ll}
\displaystyle{%
   \Omega=\Omega_0 - T \sum_{n=1}^{\infty} \langle \sigma_n(\beta)\rangle_{0\,\textrm{conn} }\,.   %
}
\end{array}
\end{equation}
Expanding the exponent in Eq.\,(\ref{EQ61}) in a series, an
arbitrary GF can be written down as
\begin{equation} \label{EQ63}
\begin{array}{ll}
\displaystyle{%
   G(1,2,\ldots L)=\sum_{n=0}^{\infty} G^{(n)}\!(1,2,\ldots L)\,.   %
}
\end{array}
\end{equation}
The contributions of the $n$-th order to the thermodynamic potential
and the GF are determined by the relations
\begin{equation} \label{EQ64}
\begin{array}{ll}
\displaystyle{%
   \langle\sigma_n(\beta)\rangle_{0\,\textrm{conn}} = \frac{g^n(-1)^n}{n!}\int_0^\beta\!\! d\tau_1\ldots  %
}\vspace{2mm}\\ %
\displaystyle{ \hspace{3mm} %
   \ldots\int_0^\beta\!\! d\tau_n \langle T_\tau\,H_C(\tau_1)\ldots H_C(\tau_n)\rangle_{0\,\textrm{conn}}\,,   %
}
\end{array}
\end{equation}
\vspace{-3mm}%
\begin{equation} \label{EQ65}
\begin{array}{ll}
\displaystyle{%
   G^{(n)}\!(1,2,\ldots L)= \frac{g^n(-1)^n\,i^L}{n!}\int_0^\beta\!\! d\tau_1'\ldots  %
}\vspace{2mm}\\ %
\displaystyle{ \hspace{0mm} %
   \ldots\int_0^\beta\!\! d\tau_n'\langle T_\tau A(1)A(2)\ldots A(L) H_C(\tau_1')\ldots H_C(\tau_n')\rangle_{0\,\textrm{conn}}.    %
}
\end{array}
\end{equation}
Owing to the properties of the normal product, $\langle\sigma_1(\beta)\rangle_{0}=0$; %
therefore, the contribution of corrections to the thermodynamic
potential in the SCF approximation arises only in the second order
of perturbation, i.e. the summation in formula (\ref{EQ62}) starts
from $n=2$. Analogously, the first-order contribution to the total
one-particle GF equals zero, so that the correction to the GF
calculated in the framework of the SCF model emerges only in the
second order of perturbation theory.

First, let us formulate the diagram technique for the field GFs. We
introduce the following graphic notations:
\begin{equation} \nonumber
\begin{array}{ll}
\displaystyle{ \hspace{0mm} %
   \scalebox{0.7}[0.7]{\includegraphics{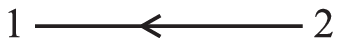}} \hspace{5mm}\textrm{for}\hspace{3mm}  G^{(0)}\!(1,2)=-\Psi^a(1)\Psi^a(2),  %
}\vspace{3mm}\\ %
\displaystyle{ \hspace{0mm} %
   \scalebox{0.7}[0.7]{\includegraphics{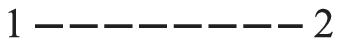}} \hspace{5mm}\textrm{for}\hspace{3mm}  \tilde{U}(1,2)=U({\bf r}_1,{\bf r}_2)\,\delta(\tau_1-\tau_2)\times  %
}\vspace{1mm}\\ %
\displaystyle{ \hspace{0mm} %
   \hspace{51mm}  \times(\delta_{\alpha_1\alpha_2}-\delta_{\alpha_1\bar{\alpha}_2}).  %
}
\end{array}
\end{equation}
Since a fermionic GF is antisymmetric with respect to the index
permutation, $G^{(0)}\!(1,2)=-G^{(0)}\!(2,1)$, in order to take this
circumstance into account, we define the direction of the Green's
line and agree that the index pointed by the arrow stands first in
the analytical record of the GF. The line, whose arrow is directed
in the opposite side, corresponds to to the same GF but with the
minus sign. Each vertex of the interaction line (dashed line) joins
two Green's lines. Since Hamiltonian (\ref{EQ46}) includes, in
pairs, operators that differ in only their isotopic indices (e.g.,
$\Psi(1')$ and $\Psi(\bar{1}')$), those two GFs include the vertex
indices which differ in a bar above.

Let us formulate the rules for calculating the contribution of the
$n$-th order to the $L$-point temperature GF:

\noindent %
1) draw $2n$ vertices linked in pairs by dashed lines;

\noindent %
2) link all the vertices with solid Green's lines in all
topologically nonequivalent ways, so that one Green's line should
enter into each vertex of the interaction line and one should leave
it;

\noindent %
3) $L/2$ vertices must be the sinks for the external Green's lines,
while the other $L/2$ vertices must be the sources for them; %
every incoming external Green's line is linked by a sequence of
solid lines to one of outgoing Green's lines; %
each such solid line is directed away from the external index, the
position of which in the analytical record of the GF is on the
right;

\noindent %
4) the vertices that are not the sinks or the sources for the
external Green's lines are linked by closed solid lines with
arbitrary directions;

\noindent %
5) no Green's line can link vertices which belong to the same dashed
line;

\noindent %
6) confront the graphic representations with their analytical
expressions: each solid line with the GF $G^{(0)}\!(1,2)$, and each
dashed line with the interaction potential $\tilde{U}(1,2)$;

\noindent %
7) the first index of the GF, which was confronted with a solid line
that enters into the vertex in question, must be taken with a bar;

\noindent %
8) carry out the integration and summation over all the variables
that were put in accordance with each vertex;

\noindent %
9) the analytical expression that has been constructed in the
indicated way according to the diagram should be multiplied by the
factor
$$ \left. g^n (-1)^{n+k+Q} \right/ 2^{k_2}\,, $$ %
where $n$ is the order of the diagram; $k$ the total number of
loops; $k_2$ the number of loops which pass through two vertices of
interaction lines;  and $Q$ the number of permutations needed for %
the external indices linked by a solid line to be arranged in the
order as they enter into the analytical record of the GF.

\begin{figure}[t!]
\centering %
\includegraphics[width = 0.99\columnwidth ]{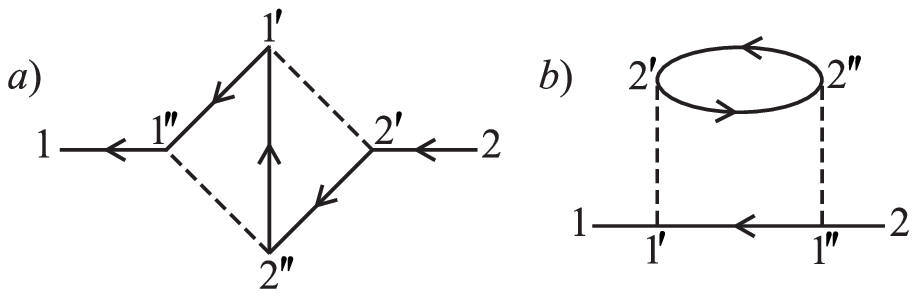} %
\caption{\label{F01} }%
\end{figure}

It is important to emphasize that, in this technique, no Green's
line can link vertices belonging to the same interaction line (item
5). Such elements are known \cite{AGD1,BKY,Mattuck} to be involved
into the diagram technique based on the ideal gas approximation.
Those diagrams can include both Green's lines leaving and entering
the same vertex (``bubbles'') and Green's lines linking the vertices
of the same interaction line (``oysters'') \cite{Mattuck}. %
The diagrams which contain such elements are impossible in the
technique proposed, because, owing to the normal form of the
perturbation Hamiltonian, there are no contractions between
operators that stand under the sign of the same $N$-product. %
The absence of such diagrams in this technique is also natural,
because the diagrams made up of these elements define the SCF
approximation which has already been taken into account in this case
in the main approximation. In Fig.\,\ref{F01}, as an example, the
diagrams of the second order are shown which define the corrections
to the one-particle GF in the field representation.

The formulated diagram technique, as well the technique based on the
approximation of non-interacting particles \cite{AGD1} admits the
diagrams to be summed up by separate blocks and the graphic methods
of summation to be used.

The diagrams for the temperature scattering matrix are constructed
by the same rules, as those for the construction of GFs. The former
differ from the latter by the absence of external lines. As is known
\cite{AGD1,BKY}, it does not allow the graphic summation of the
infinite sequences of diagrams to be carried out in this case.

For the practical use of the diagram technique, it is more
convenient to pass to the frequency representation. In this case,
the rules of the diagram technique undergo the following
modifications:

\noindent %
1) every Green's line is associated with Fourier-component
(\ref{EQ57}) $G^{(0)}\!(1,2;\omega_n)$, and every external incoming
line should be associated with a frequency with the minus sign;

\noindent %
2) every dashed line is associated with the potential
$\underline{\tilde{U}}(1,2)=U({\bf r}_1,{\bf
r}_2)(\delta_{\alpha_1\alpha_2}-\delta_{\alpha_1\bar{\alpha}_2})$;

\noindent %
3) the frequency conservation law must be fulfilled: the sum of the
frequencies of Green's lines which enter the end points of every
dashed line of the interaction is equal to the sum of frequencies of
the outgoing lines, which is taken into account by introducing the
multiplier $\Delta(\omega_{n_1}+\omega_{n_2}-\omega_{n_3}-\omega_{n_4})$; %
with $\Delta(\omega)=1$ if $\omega=0$, and $\Delta(\omega)=0$ otherwise; %

\noindent %
4) the additional multiplier $T^{n-L/2}$ emerges before the
expression that correspond to the diagram.

\begin{figure}[t!]
\centering %
\includegraphics[]{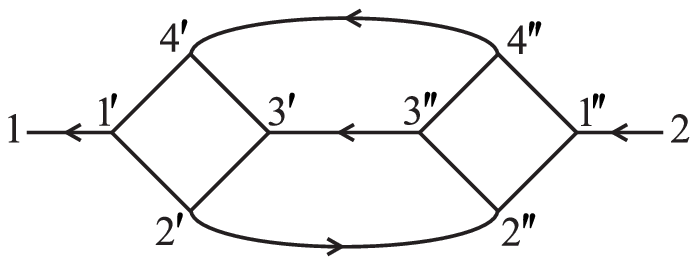}
\caption{\label{F02} }%
\end{figure}

Now, let us formulate the rules of the diagram technique in the case
where the quasiparticle description is used. Taking into account
that the adjacent operators can be permutated under the sign of
normal product with changing the sign, Hamiltonian (\ref{EQ47}) can
be represented in the form
\begin{equation} \label{EQ66}
\begin{array}{ll}
\displaystyle{%
   H_C= \frac{1}{4!}\sum_{1234} \tilde{U}_{1234} N(\gamma_1\gamma_2\gamma_3\gamma_4)\,,  %
}
\end{array}
\end{equation}
where the antisymmetrized potential
\begin{equation} \label{EQ67}
\begin{array}{ll}
\displaystyle{%
   \tilde{U}_{1234}=U_{1234}\!+\!U_{1342}\!+\!U_{1423}\!+\!U_{2314}\!+\!U_{2431}\!+\!U_{3412}-  %
}\vspace{3mm}\\ %
\displaystyle{ \hspace{3mm} %
  -U_{1243}\!-\!U_{1324}\!-\!U_{1432}\!-\!U_{2341}\!-\!U_{2413}\!-\!U_{3421}.   %
}
\end{array}
\end{equation}
was introduced. Let us introduce a notation for the potential
\begin{equation} \label{EQ68}
\begin{array}{ll}
\displaystyle{%
   \tilde{U}(1,2,3,4)=\tilde{U}_{1234}\,\delta(\tau_1-\tau_2)\,\delta(\tau_1-\tau_3)\,\delta(\tau_1-\tau_4)\,,  %
}
\end{array}
\end{equation}
which is antisymmetric with respect to the permutation of variables.
A single digit in the potential $\tilde{U}(1,2,3,4)$ designates the
following set of variables: $1\!=\!(i_1,\alpha_1,\tau_1)$. Every
square in the diagram is confronted with the matrix element of the
interaction potential %
\vspace{0mm}%
\begin{equation} \nonumber
\begin{array}{ll}
\displaystyle{%
   \tilde{U}(1,2,3,4)  \hspace{8mm}\textrm{---}\hspace{8mm} \scalebox{0.65}[0.65]{\includegraphics[bb = 28 482 85 510]{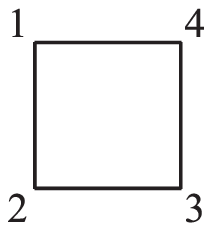}} \quad  ,  
}\vspace{3mm} %
\end{array}
\end{equation}
while every Green's line with the zero-order approximation of the GF
in the quasiparticle representation. Indices at the square vertices
must be arranged either clockwise or counterclockwise in that order
as they enter into the expression for the matrix element (\ref{EQ68}). %

Thus, the rules for calculating the contribution of the $n$-th order
to the Green's quasiparticle function are as follows:

\noindent %
1) draw $n$ squares that correspond to the matrix elements of the
interaction potential;

\noindent %
2) link the vertices of different squares with solid lines in all
such topologically nonequivalents ways, that two vertices of a
square serve as sinks for one Green's line each and two other
vertices of the same square be the sources of one Green's line each;

\noindent %
3) confront the graphic representations with their analytical
expressions; sum up over the variables related to each vertex and
integrate over ``time'' variables;

\noindent %
4) put the multiplier $(-1)^{n+k+Q}/P_n$ before the expression that
has been constructed according to the diagram, where $n$ is the
order of the diagram, $k$ the number of closed loops in it, $Q$ the
number of permutations which are necessary to arrange the external
indices linked by a solid line in that order as they enter into the
analytical record of the GF, $P_n$ the number of permutations of the
indices in the square vertices that do not result in new expressions. %

In the quasiparticle representation, we get only one diagram of the
second order which is shown in Fig.\,\ref{F02}.

For the frequency representation of quasiparticle GFs, the rules are
modified in the same way as in the case of field GFs.

{\bf 8.} Similarly as in the quantum-field approach that uses the
model of independent particles as the zero-order approximation, the
concept of self-energy and vertex parts can be introduced in the
approach that is developed here, and the Dyson's equation that
couples those two functions can be derived. The equation for a
one-particle GF can be presented in the following form:
\begin{equation} \label{EQ69}
\begin{array}{ll}
\displaystyle{%
   \frac{\partial G(1,2)}{\partial\tau_1}=-\delta(1-\bar{2})-  %
}\vspace{2mm}\\ %
\displaystyle{ \hspace{3mm} %
  -\int\! d3\,\big[H(1,\bar{3})+W(1,\bar{3})+\Sigma(1,\bar{3})\big]G(3,2)\,,   %
}
\end{array}
\end{equation}
where the self-energy function $\Sigma(1,2)$ is defined by the
relation
\begin{equation} \label{EQ70}
\begin{array}{ll}
\displaystyle{%
   \int\! d3\,\Sigma(1,\bar{3})\,G(3,2)=\frac{1}{2}\int\! d3\,\tilde{U}(1,3)\times   %
}\vspace{3mm}\\ %
\displaystyle{ \hspace{2mm} %
  \times\big[ G(1,3,\bar{3},2)\!+\!G^{(0)}\!(\bar{3},3)\,G(1,2)\!-\!2\,G^{(0)}\!(\bar{3},1)\,G(3,2)\big].    %
}
\end{array}
\end{equation}
Formula (\ref{EQ69}) includes the functions
\begin{equation} \label{EQ71}
\begin{array}{ll}
\displaystyle{%
   H(1,\bar{3})=H_{\alpha_1\bar{\alpha}_3}(q_1,q_3)\,\delta(\tau_1-\tau_3)\,,  %
}\vspace{3mm}\\ %
\displaystyle{ \hspace{0mm} %
   W(1,\bar{3})=W_{\alpha_1\bar{\alpha}_3}(q_1,q_3)\,\delta(\tau_1-\tau_3)\,,    %
}
\end{array}
\end{equation}
where
\begin{equation} \nonumber
\begin{array}{ll}
\displaystyle{%
   H_{\alpha\alpha'}(q,q')= %
   \left[
     \begin{array}{cc}
       0          & H(q,q') \\
       -H^*(q,q') & 0       \\
     \end{array}
   \right]\,, %
}\vspace{4mm}\\ %
W_{\alpha\alpha'}(q,q')= %
   \left[
     \begin{array}{cc}
       \Delta(q,q') &  W(q,q')         \\
       -W^*(q,q')   &  -\Delta^*(q,q') \\
     \end{array}
   \right]\,. %
\end{array}
\end{equation}
With regard for the equations for a GF in the SCF approximation, we
obtain the known relation for the self-energy function:
\begin{equation} \label{EQ72}
\begin{array}{ll}
\displaystyle{%
   G(1,2)=G^{(0)}\!(1,2) + \int\!\! d3 d4\, G(1,\bar{3})\,\Sigma(3,\bar{4})\,G^{(0)}\!(4,2)\,. %
}
\end{array}
\end{equation}

The vertex part $\Gamma(1,2,3,4)$ is defined by the formula
\begin{equation} \label{EQ73}
\begin{array}{ll}
\displaystyle{%
   G(1,2,3,4)=   %
}\vspace{3mm}\\ %
\displaystyle{ \hspace{3mm} %
  = G(1,2)\,G(3,4) + G(1,4)\,G(2,3) - G(1,3)\,G(2,4)+     %
}\vspace{3mm}\\ %
\displaystyle{ \hspace{2mm} %
  + \int\!\! d1' d2' d3' d4'\,\Gamma(1',2',3',4')\times    %
}\vspace{3mm}\\ %
\displaystyle{ \hspace{2mm} %
  \times\, G(\bar{1}',1)\,G(\bar{2}',2)\,G(\bar{3}',3)\,G(\bar{4}',4)\,.   %
}
\end{array}
\end{equation}

From Eq.\,(\ref{EQ70}), taking Eq.\,(\ref{EQ73}) into account, we
obtain the Dyson's equation which establishes a relation between the
self-energy and vertex parts in Fermi systems with spontaneously
broken symmetry:
\begin{equation} \label{EQ74}
\begin{array}{ll}
\displaystyle{%
   \Sigma(1,2)\!=\!\frac{1}{2}\int\! d3\,\tilde{U}(1,3)\Big\{\!\big[ G(3,\bar{3})-G^{(0)}\!(3,\bar{3}) \big]\delta(1-\bar{2})-  %
}\vspace{3mm}\\ %
\displaystyle{ \hspace{3mm} %
  - 2 \big[ G(3,1)-G^{(0)}\!(3,1)\big] \delta(2-3)\Big\} +     %
}\vspace{3mm}\\ %
\displaystyle{ \hspace{3mm} %
  + \frac{1}{2}\int\! d1' d2' d3' d3\,\tilde{U}(1,3)\times    %
}\vspace{3mm}\\ %
\displaystyle{ \hspace{3mm} %
  \times\Gamma(1',2',3',2)\,G(\bar{1}',1)\,G(\bar{2}',3)\,G(\bar{3}',3)\,.   %
}
\end{array}
\end{equation}
The poles of the vertex function introduced by relation (\ref{EQ73})
define the dispersion law of collective excitations in the many-body
system.

The methods of quantum field theory applied in statistical physics
were extended in this work to describe any eligible states in
non-relativistic Fermi systems with spontaneously broken symmetry.
We managed to do this, mainly, owing to two circumstances. First, it
is the SCF model formulated in the most general form that was used
as the main approximation. Secondly, it is the procedure for
calculating the quasiaverages which uses the fields that are defined
by this model. It is essential that, in the given approach, the
correlation Hamiltonian considered as a perturbation can be
presented in the normal form, which allows a plenty of diagrams not
contributing to the final result to be excluded from consideration
and the diagram technique to be presented in a compact form. %
The approach suggested does not contain any assumptions and is based
only on the general principles of quantum mechanics and statistical
physics. It can be applied for regular researches of equilibrium
properties of many-particle systems with spontaneously broken
symmetry (magnetically and spatially ordered, superconducting,
superfluid, etc. systems) and the phenomena in them at a microscopic
level. The method proposed can be extended onto the description of
non-relativistic Bose systems with spontaneously broken symmetry, in
particular, of superfluid systems with broken phase symmetry
\cite{Poluektov5,Poluektov6}. In author's opinion, the general
approach developed in this work can also be effectively used for a
proper description of states with spontaneously broken symmetry in
the relativistic field theory and the theory of elementary particles
\cite{NL,Goldstone}.

\end{document}